\documentclass[a4paper]{llncs}

\usepackage[utf8x]{inputenc}

\usepackage{amsmath}

\usepackage{amsthm}
\usepackage{amssymb}
\usepackage{breqn}
\usepackage{framed}
\usepackage{array}
\usepackage{xfrac}
\usepackage{tikz}
\usepackage{tkz-berge}
\usepackage{enumerate}
\usepackage{float}
\usepackage{bookmark}
\usepackage{hyperref}
\hypersetup{
  pdftitle = {Graph algorithms},
  pdfauthor = {},
  pdfkeywords = {},
  pdfstartview=FitH,
  unicode=true
}

\usepackage{hypcap}

\newcommand{\ket}[1]{\left| #1 \right\rangle}

\newcommand{\braket}[2]{\left\langle #1 | #2 \right\rangle}

\newtheorem{algorithm}{Algorithm}

\usetikzlibrary{calc}

\begin{document}

\title{Span-program-based quantum algorithms for graph bipartiteness and connectivity
\thanks{%
This work has been supported by the \textsc{ERC Advanced Grant} Methods for Quantum Computing.
}
}
\author{\texorpdfstring{Agnis Āriņš}{Agnis Āriņš}}

\institute{University of Latvia, Raiņa bulvāris 19, Riga, LV-1586, Latvia}

\maketitle

\begin{abstract}
Span program is a linear-algebraic model of computation which can be used to design quantum algorithms. 
For any Boolean function there exists a span program that leads to a quantum algorithm with optimal quantum query complexity. In general, finding such span programs is not an easy task.

In this work, given a query access to the adjacency matrix of a simple graph $G$ with $n$ vertices, we provide two new span-program-based quantum algorithms:
\begin{itemize}
\item an algorithm for testing if the graph is bipartite that uses $O(n\sqrt{n})$ quantum queries;
\item an algorithm for testing if the graph is connected that uses $O(n\sqrt{n})$ quantum queries.
\end{itemize}
\end{abstract}

\section{Introduction}

The concept of a span program as a linear-algebraic model of computation is not new. It was introduced by Karchmer and Wigderson in 1993 \cite{KW93} and has many applications in classical complexity theory. Span programs can be used to evaluate  decision problems.
In 2008 Reichardt and Spalek \cite{Rei08} introduced a new complexity measure for span programs -- witness size, which, as Reichardt showed later in \cite{Rei09,Rei11}, has strong connection with the quantum query complexity. There is a quantum algorithm for evaluating span programs \cite{Rei09} and these two complexity measures are essentially equivalent. The difficulty is to come up with a span program with a good witness size complexity.  

In \cite{Rei08} authors dealt with bounded-size  span programs  evaluating  Boolean functions  each on $O(1)$  bits and posed an open question -- do there exist interesting quantum algorithms based directly on asymptotically large span program? Belovs used span programs to construct learning graphs \cite{Bel12,Bel12a}. He also used span program approach for the matrix rank problem \cite{Bel11}. In \cite{Amb13} Ambainis et al. came up with a simple yet powerful span program for the graph collision problem.

In this paper, we extend the family of algorithms based on span programs. We present two new span-program-based quantum algorithms -- an $O(n\sqrt{n})$ algorithm for the graph bipartiteness problem and an $O(n\sqrt{n})$ algorithm for the graph connectivity problem. Both algorithms in the quantum query sense are optimal because the witness sizes match the quantum query complexity lower bounds \cite{Berz04,Dur04} for these problems. Thus we demonstrate that span programs can be useful also for the problems with an asymptotically large input and possibly our algorithms could be building blocks for bigger span programs in the future.

The graph connectivity problem has been studied before \cite{Dur04} and there already exists a $O(n\sqrt{n})$ quantum query algorithm which requires $O(n)$ qubits of quantum memory. The advantage of our algorithm is that it uses only $O(\log n)$ qubits of quantum memory because the span program $P_2$ uses a vector space with $O(n^2)$ dimensions. Similarly for the graph bipartiteness problem. It can be solved with the breadth-first search method \cite{Furr08} which uses $O(n)$ qubits of quantum memory, but our approach with a span program requires $O(\log n)$ qubits of quantum memory.

\section{Preliminaries}

In this paper, we present algorithms which work on simple graphs, given in adjacency model. If the given graph has $n$ vertices then the input size for an algorithm is $n \times n$ and we assume that the input variable $x_{i,j}$ corresponds to the value of entry in i-th row and j-th column of the adjacency matrix.

\subsection{Span programs}

\begin{definition}[\cite{Amb13}]
A \emph{span program} $P$ is a tuple $P=\left(H, \ket{t}, V\right)$, where $H$ is a finite-dimensional Hilbert space, $\ket{t}\in H$ is called the target vector, and $V=\{V_{i,b}|i \in [n], b \in \{0,1\}\}$, where each $V_{i,b} \subseteq H$ is a finite set of vectors.

Denote by $V(x) = \bigcup{\{V_{i,b}|i\in [n], x_i=b\}}$. The span program is said to \emph{compute} function $f:D \rightarrow \{0,1\}$, where the domain $D \subseteq \{0,1\}^n$, if for all $x\in D$, 
   \[f(x)=1 \iff \ket{t} \in \operatorname{span}(V(x)).\]
\end{definition}

Basically, what this definition says is that for each input variable $x_i$ we have two sets of vectors (as the span program authors, we define these vectors in advance) -- $V_{i,0}$ and $V_{i,1}$. If $x_i=b$ then we say that vectors from the set $V_{i,b}$ are available and vectors from the set $V_{i,1-b}$ are not available. If some vector is included in both sets $V_{i,0}$ and $V_{i,1}$ then we say that it is a free vector -- it is always available.

The function returns $1$ iff the target vector can be expressed as a linear combination of the available vectors, otherwise it returns $0$.

\begin{definition}[\cite{Amb13}]
\begin{enumerate}[(1)]
\item A \emph{positive witness} for $x\in f^{-1}(1)$ is a vector $w=(w_v),v\in V(x)$, such that $\ket{t}=\sum_{v\in V(x)}{w_v v}$. The \emph{positive witness size} is 
\[\operatorname{wsize}_1(P):=\max_{x\in f^{-1}(1)}{\min_{w:\text{witness of }x}{\|w\|^2}}.\]

\item A \emph{negative witness} for $x\in f^{-1}(0)$ is a vector $w\in H$, such that $\braket{t}{w}=1$ and for all $v\in V(x)$: $\braket{v}{w}=0$. The \emph{negative witness size} is \[\operatorname{wsize}_0(P):=\max_{x\in f^{-1}(0)}{\min_{w:\text{witness of }x}{\sum_{v\in V}{{\braket{v}{w}}^2}}}.\]

\item The \emph{witness size of a program} $P$ is
\[\operatorname{wsize}(P):=\sqrt{\operatorname{wsize}_0(P)\cdot \operatorname{wsize}_1(P)}.\]

\item The \emph{witness size of a function} $f$ denoted by $\operatorname{wsize}(f)$ is the minimum witness size of a span program that computes $f$.
\end{enumerate}
\end{definition}

\begin{theorem}[\cite{Rei09,Rei11}]
\label{thm:span}
$Q(f)$ and $\operatorname{wsize}(f)$ coincide up to a constant factor. That is, there exists a constant $c > 1$ which does not depend on $n$ or $f$ such that $\frac{1}{c}\operatorname{wsize}(f) \leq Q(f) \leq c \cdot \operatorname{wsize}(f)$.
\end{theorem}

\section{Span program for testing graph bipartiteness}

A bipartite graph is a graph whose vertices can be divided into two disjoint sets
such that there is no edge that connects vertices of the same set. An undirected graph is bipartite iff it has no odd cycles.

\begin{algorithm}
There exists a span program $P$ which for a graph with $n$ vertices detects if the graph is bipartite with $wsize(P) = O(n\sqrt{n})$.
\end{algorithm}

\begin{proof}
We will make a span program which detects if a graph has an odd cycle.

Let $n=|G|$ be a number of vertices in the given graph $G$. Then the span program is as follows:

\begin{framed}
\begin{center}Span program $P_1$ for testing graph bipartiteness
\end{center}
\begin{itemize}
\item $H$ is a $(2n^2+1)$ dimensional vector space with basis vectors $\{\ket{0}\}\cup\{\ket{v_{k,b}} | v,k \in [1..n], b\in\{0,1\}\}$.\par
\item The target vector is $\ket{0}$.\par
\item For every $k\in[1..n]$ make available the free vector $\ket{0}+\ket{k_{k,0}}+\ket{k_{k,1}}$.\par
\item For every $k\in[1..n]$, for every edge $u-v$ (where input $x_{u,v}=1$), make available the vectors $\ket{u_{k,0}}+\ket{v_{k,1}}$ and $\ket{u_{k,1}}+\ket{v_{k,0}}$.\par
\end{itemize}
\end{framed}

The states in the span program $P_1$ are mostly in the form $\ket{v_{k,b}}$ where $v$ is vertex index, $k$ represents vertex from which we started our search for an odd length cycle and $b$ represents the parity of the current path length. The first subindex $k$ in state $\ket{v_{k,b}}$ can also be considered as the subspace index for the subspace $V_k = \operatorname{span}(\{\ket{v_{k,b}} | v \in [1..n], b\in\{0,1\}\})$. Vectors corresponding to edges are in the form $\ket{u_{k,b}}+\ket{v_{k,1-b}}$ consisting from sum of two states which both belong to same subspace $V_k$.

In the span program $P_1$ the target vector $\ket{0}$ can only be expressed as a linear combination of the available vectors if at least one of the vectors in the form $\ket{k_{k,0}}+\ket{k_{k,1}}$ can be expressed. Without loss of generality, if there is an odd length cycle $v_1-v_2-\dots-v_{(2j+1)}-v_1$ then the target vector can be expressed by taking the
vectors corresponding to the edges of this cycle, alternatingly with plus and minus sign $$\ket{0}=\left(\ket{0}+\ket{1_{1,0}}+\ket{1_{1,1}}\right)-\left(\ket{1_{1,0}}+\ket{2_{1,1}}\right)+\dots-\left(\ket{(2j+1)_{1,0}}+\ket{1_{1,1}}\right)$$
therefore the span program $P_1$ will always return $1$ when the given graph is not bipartite. 

From the other side, if there is no odd length cycle then none of the vectors in the form of $\ket{k_{k,0}}+\ket{k_{k,1}}$ can be expressed using the available vectors from $P_1$.
To cancel out the state $\ket{k_{k,0}}$ we should be using a vector $\ket{k_{k,0}}+\ket{v_{k,1}}$ corresponding to some edge $k-v$ where $v$ is some vertex adjacent to $k$ because no other vector contains the state $\ket{k_{k,0}}$. By doing so we move from the state $\ket{k_{k,0}}$ to the state $\ket{v_{k,1}}$ (possibly with some coefficient other than $1$) which has the parity bit flipped. Similarly, to cancel out the state $\ket{v_{k,1}}$ we should be using a vector corresponding to some edge going out from vertex $v$. To stop this process we need to reach the state $\ket{k_{k,1}}$. It can be done only if there is an odd cycle because the path must be closed and the parity bit restricts it to odd length. When there is no odd cycle, span program $P_1$ will always return $0$. 

We can conclude that $P_1$ indeed computes the expected function. It remains to calculate the witness size of $P_1$.

For the case when there is an odd cycle we need to calculate the positive witness size.
If there is an odd cycle $v_1-v_2-\dots-v_{d}-v_1$ with length $d$ then the target vector can be expressed in this way $$\ket{0}=1\cdot\left(\ket{0}+\ket{1_{1,0}}+\ket{1_{1,1}}\right)+(-1)\cdot\left(\ket{1_{1,0}}+\ket{2_{1,1}}\right)+\dots+(-1)\cdot\left(\ket{d_{1,0}}+\ket{1_{1,1}}\right)$$
and the positive witness $w$  here consists only from $d+1$ entries $\pm1$ therefore  $\|w\|^2=d+1$.

If $v_1-v_2-\dots-v_{d}-v_1$ is a cycle then also $v_2-v_3-\dots-v_{d}-v_1-v_2$ is a cycle and therefore the target vector can also be expressed in this way  $$\ket{0}=\left(\ket{0}+\ket{2_{2,0}}+\ket{2_{2,1}}\right)-\left(\ket{2_{2,0}}+\ket{3_{2,1}}\right)+\dots-\left(\ket{1_{2,0}}+\ket{2_{2,1}}\right)$$
the same follows for all $d$ vertices in this cycle and the target vector therefore can be expressed in atleast $d$ different ways. We can combine these $d$ ways each taken with coefficient $1/d$ and then we get that the positive witness size
\begin{equation}
\operatorname{wsize}_1(P_1) \leq d*(1/d)^2*(d+1)<2
\end{equation}
To estimate the negative witness size we must find a negative witness $w'$. We derive $w'$ by defining how it acts on basis vectors. From definition ${\braket{w'}{0}}=1$. For every $k$ we must have $\braket{w'}{\left(\ket{0}+\ket{k_{k,0}}+\ket{k_{k,1}}\right)}=0$ therefore lets pick $w'$ in such a way that $\braket{w'}{k_{k,0}}=0$ and $\braket{w'}{k_{k,1}}=-1$. Now repeat the following steps until no changes happen: 
\begin{itemize}
\item for every available vector $\ket{u_{k,0}}+\ket{v_{k,1}}$ if $\braket{w'}{u_{k,0}}$ is defined then define $\braket{w'}{v_{k,1}}=-\braket{w'}{u_{k,0}}$.
\item for every available vector $\ket{u_{k,1}}+\ket{v_{k,0}}$ if $\braket{w'}{u_{k,1}}$ is defined then define $\braket{w'}{v_{k,0}}=-\braket{w'}{u_{k,1}}$.
\end{itemize}
For all not yet defined $\braket{w'}{v_{k,b}}$ define $\braket{w'}{v_{k,b}}=0$.

For any given vector $v$ in span program $P_1$ the value ${\braket{w'}{v}}^2\le1$. 
The total number of vectors does not exceed $n+n^3$ therefore the negative witness size is 
\begin{equation}
\operatorname{wsize}_0(P_1) \leq 1\cdot(n+n^3)
\end{equation}

Combining positive and negative witness sizes we obtain the upper bound for witness size which also corresponds to quantum query complexity  
\begin{equation}
\operatorname{wsize}(P_1) =  \sqrt{\operatorname{wsize}_0(P_1)\cdot \operatorname{wsize}_1(P_1)} = O\left(n\sqrt{n}\right)
\end{equation}

\end{proof}
\noindent

\section{Span program for testing graph connectivity}

A graph is said to be connected if every pair of vertices in the graph is connected. If in an undirected graph one vertex is connected to all other vertices then by transitivity the graph is connected.

\begin{algorithm}
There exists a span program $P$ which for a graph with $n$ vertices detects if the graph is connected with $wsize(P) = O(n\sqrt{n})$.
\end{algorithm}

\begin{proof}

Let $n=|G|$ be a number of vertices in the given graph $G$. Then the span program is as follows:

\begin{framed}
\begin{center}Span program $P_2$ for testing graph connectivity
\end{center}
\begin{itemize}
\item $H$ is a $n^2-1$ dimensional vector space with basis vectors $\{\ket{v_k} | v \in [0..n], k \in [2..n]\}$.\par
\item The target vector is $\ket{t} = \ket{0_2}+\ket{0_3}+\dots+\ket{0_n}$.\par
\item For every $k\in[2..n]$ make available the free vector $\ket{0_k}+\ket{1_k}-\ket{k_k}$.\par
\item For every $k\in[2..n]$, for every edge $u-v$ (where $u,v\in[1..n]$ and input $x_{u,v}=1$), make available the vector $\ket{u_k}-\ket{v_k}$.\par
\end{itemize}
\end{framed}

If all vertices are reachable from vertex with index $1$ then the given graph is connected. Here we use Belov's \cite{Bel12b} span program for \emph{s-t connectivity} as subroutine. This subroutine checks if in a given graph there is a path from the vertex $s$ to the vertex $t$. The span program for it has the target vector $\ket{s}-\ket{t}$ and for each edge $i-j$ (input $x_{i,j}=1$) we can use the vector $\ket{i}-\ket{j}$.

In span program $P_2$, by using this subroutine $n-1$ times, we check if all other vertices are connected to vertex with index $1$. We create a separate subspace $V_k = \operatorname{span}(\{\ket{v_{k}} | v \in [0..n]\})$ for each such subroutine call to avoid any interference between them, which is a common technique \cite{Rei09} how to compose span programs. The span program returns $1$ when all vertices are connected, but otherwise it returns $0$.

For the case when the given graph is connected we need to calculate the positive witness size. In each \emph{s-t} subroutine the shortest path length from the vertex $s$ to the vertex $t$ can not be larger than $n-1$. Threfore each vector from the set $\left\{\ket{0_k} | k\in[2..n]\right\}$ requires no more than $n$ vectors to express it. There are $n-1$ such subroutines. The positive witness size is 
\begin{equation}
wsize_1(P_2)\leq n\cdot(n-1) \leq n^2
\end{equation}

To estimate the negative witness size we must find a negative witness $w'$. We derive $w'$ by defining how it acts on the basis vectors. From definition ${\braket{w'}{t}}=1$. We need to talk about negative witness only when some vertex $v$ is not connected to vertex with index 1. Then the vertex $v$ belongs to different connected component than vertex with index 1.  Lets name this connected component $C_v$ and let the count of vertices in this connected component be $d_v$. Lets pick $w'$ in such a way that for each vertex ${v_k}\in{C_v}$ set
$\braket{w'}{0_k}=1/{d_v}$ 
and  for  each vertex
${v_z}\notin{C_v}$ set
$\braket{w'}{0_z}=0$. 

For $k\in[2..n]$ we must have $\braket{w'}{(\ket{0_k}+\ket{1_k}-\ket{k_k})}=0$ therefore set $\braket{w'}{1_k}=-\braket{w'}{0_k}$ and $\braket{w'}{k_k}=0$. Now repeat the following step until no changes happen: 
for every available vector $\ket{u_k}-\ket{v_k}$ if $\braket{w'}{u_k}$ is defined then define $\braket{w'}{v_k}=\braket{w'}{u_k}$.
For all other not yet defined basis vectors $\ket{v_k}$ set $\braket{w'}{v_k}=0$. 

With such negative witness $w'$ choice the overall negative witness size will only get increased by vectors which correspond to nonexistent graph edges which connects $C_v$ with other connected components in graph - i.e. border edges. 
An edge $u-v$ is a border edge if $u\in C_v$ and $v\notin C_v$. 
To a border edge $u-v$ correspond the vectors  $\ket{u_k}-\ket{v_k}$ where $k\in[2..n]$ but only $d_v$ from these vectors will have $\braket{w'}{u_k} \neq \braket{w'}{v_k}$ and each such vector increases the negative witness size by value $(1/d_v)^2$. For $C_v$ there are at most $d_v\cdot(n-1)$ border edges therefore the overall negative witness size is

\begin{equation}
\operatorname{wsize}_0(P_2) \leq d_v^2\cdot(n-1)\cdot(1/d_v)^2 \leq n
\end{equation}
Combining the positive and negative witness sizes we obtain the upper bound for the witness size which also corresponds to the quantum query complexity  
\begin{equation}
\operatorname{wsize}(P_2) =  \sqrt{\operatorname{wsize}_0(P_2)\cdot \operatorname{wsize}_1(P_2)} = O\left(n\sqrt{n}\right)
\end{equation}

\end{proof}

\section*{Acknowledgements}

I am grateful to Andris Ambainis for the suggestion to solve the graph problems with span programs, and for many useful comments during the development of the paper.

\bibliographystyle{splncs03}

\phantomsection
\addcontentsline{toc}{chapter}{References}
\bibliography{quantum,tw}

\end{document}